# Robust Packaged Fiber-Microcavity Device with over One Billion Q-factor

**Fangxing Zhang[1], Shengnan HuangFu[1], Jialve Sun[1], Shengqiang Ji[1], Yanjie Bai[1], Yunfeng Xiao[1,2]**

[1]Peking University Yangtze Delta Institute of Optoelectronics, Nantong, Jiangsu, 226010, China
[2]State Key Lab for Mesoscopic Physics and Frontiers Science Center for Nano-optoelectronics, School of Physics, Peking University, Beijing,100871, China

*Abstract*—Whispering gallery mode (WGM) microcavities can confine photons within a microscale volume for long periods of time, strongly enhancing light-matter interactions, and making it a crucial platform in optical science and applications. Current research on microcavity coupling system relays on precise mechanical coupling with microscope monitoring, and its resonance properties are extremely sensitive to external interference, which greatly limits the practical application of microcavities. Therefore, a novel packaged fiber-microcavity device with air/water tightness and stable temperature characteristics is proposed in this paper. A variety of fixatives with different Young's modulus gradients and low coefficients of thermal expansion are used to design a package structure with overall vibration isolation and buffering effect, in order to ensure the stability of the transmission spectrum and improve the robustness of the microcavity module. Through the performance characterization test of the device, it is proved that the packaged microcavities can maintain a quality factor as high as $10^9$, and has the advantages of compact size, strong robustness, and versatility. This research work is of great significance to promote the large-scale application of WGMs in high-speed optical communication, nonlinear optics, narrow linewidth lasers, and ultra-high sensitivity sensing.

*Index Terms*-whispering gallery mode, optical microcavities, quality factors, packaging method, robustness

## I. INTRODUCTION

Resonator is one of the essential basic components in the optical system, and also the structural basis of the optical oscillator, filter, laser, etc. Whispering gallery mode (WGM) optical microcavity is a new resonator with ultra-high quality factor (Q factor) and small mode volume[1, 2]. After decades of development, it has become an important platform for research on soliton optical frequency combs [3, 4], narrow linewidth lasers [5, 6], coherent control [7, 8], and precision sensing [9-11]. However, the optical microcavity system is highly sensitive to the surroundings outside the resonator. First, the condensation of water and deposition of microdust particles on the microcavity surface when it is exposed to the air, resulting in the enhanced loss of the cavity surface, which will significantly reduce the Q factor. Second, current research on microresonators relays on precise mechanical coupling with microscope monitoring, and the resonant property is susceptible to the ambient vibration, temperature, and humidity changes. These can both limit the development and application of microcavity-based practical devices. Therefore, high Q factor microresonators with stable mode and compact structure obtained through reasonable packaging technology, are the prerequisite for the actual large-scale application of optical microcavities.

Over the last decade, to find a general microcavity packaging method, a variety of studies have been reported. Spot-packaged method [12,13] and full-encapsulated packaging [14, 15] are the microcavity packaging technologies earlier proposed, which use low-refractive-index polymers as encapsulation materials to fix the coupling region and encapsulate the whole coupling system respectively. In view of the high requirements of the dispensing process for the spot-packaging method, and fully encapsulated not convenient for direct detection with evanescent field, Dong Yong-Chao et al. [16] proposed an ingenious packaged microsphere-taper coupling system which consists of a glass tube and two glass plates with a high Q-factor of $1.08 \times 10^8$ in 2015. The maintenance of Q and a stable spectrum were realized by placing the packaged structure in a sealed organic glass box. In 2022, Wang Meng-Yu et al. [17] proposed a novel indirect packaged optical microcavity device and the Q factor reached $5.1 \times 10^7$. The resonant mode was in a relatively stable state for up to 1000s. In addition, in order to further explore the coupling packaging technology of microcavities and its application prospects, Zhao Guang-Ming et al. [18] reported Raman lasing and the optical analog of electromagnetically-induced-transparency (EIT) in a WGM microtoroid resonator embedded in a low-refractive-index polymer matrix together with a tapered fiber coupler and achieved high Q factors and critical coupling. Before long, Zhang Meng et al. [19] reported a creative in-fiber microsphere resonator-based integrated device which provides greatly improved mechanical stability. Yang Yu et al. [20] designed a packaged platform that can steadily seal the resonator and tapered fiber in a nitrogen atmosphere and achieved up to the fifth-order cascaded stimulated Brillouin scattering (SBS) light in a $CaF_2$ micro-disk resonator. To date, studies on high Q factor microcavity packaged devices have yielded relatively fruitful achievements, but there are few reports on long-term stability tracking experiments of the Q factors and resonant properties after packaging. How to effectively encapsulate optical microcavities remains a huge challenge for researchers.

In this letter, we propose an ultra-high Q factor packaged silica microresonator device coupled to a tapered fiber. The packaged structure has excellent air tightness, water tightness, vibration isolation, and thermal properties. Without changing the adjusted coupling state, the Q factors up to $10^9$ are achieved with good parameter consistency (Q factor error < 10%, FSR error < 2%). Furthermore, the durability and vibration tests of the packaged microresonators are done to verify the robustness of the proposed packaging method. This study has vital application value in promoting the development of WGM-based practical devices, such as narrow linewidth lasers, ultra-narrow optical filters, and soliton light sources.

## II. FABRICATION AND PACKAGING TECHNIQUE

In experiments, the microresonators are fabricated during an automated process, which is by irradiating $CO_2$ lasers on dehydroxylated high-purity fused silica microrods that are mounted on a rotating spindle. After the geometries of the microresonators are defined, a reflow process is introduced to refine surface uniformity. For the sake of obtaining maximum power coupling efficiency and ensuring high Q resonances and long-term stable transmission spectra, the fiber taper is fabricated by using the method of heat-and-pull with low insertion loss, and based on this, we design an innovative structure for packaging the coupling system.

The schematic diagram of the microrod-taper coupling packaged structure is shown in Fig. 1 (a). The core part is a specially machined envelope made of aluminum alloy which is lined with quartz substrate and has two mutually perpendicular grooves used to place and fix the silica microrod and tapered fiber respectively. Circular holes are left at both ends of the shell to make the fiber pass through and facilitate the subsequent standardized end-to-end operation. O-rings and silicone rubber seals are used to isolate water vapor and atmospheric particles, so as to ensure the excellent air and water tightness of the encapsulated microresonators. Compared with the traditional UV curing adhesive or high-temperature curing method, we select multilayer fixatives with different Young's modulus gradients and low thermal expansion coefficients to fix the resonator and coupling waveguide. This packaging method makes the WGM microcavities more immune to temperature fluctuation, external vibration, and airflow. On this basis, the Tec temperature control system and the standardized APC polarization-maintaining fiber connector are adopted to prevent the resonant position shift and improve the coherent signal-to-noise ratio and spectrum stability of the system respectively. Fig. 1 (b) shows the image of one optical microresonator after packaging. This packaging scheme is a successful attempt, which avoids the problem of air leakage in the traditional process and the problem of coupling point drift in long-term use, and greatly improves the stability and service life of the microresonators.

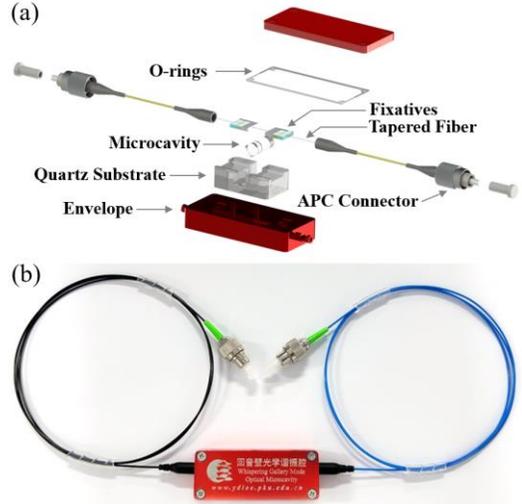

**Fig. 1.** (a) Schematic diagram of the microrod-taper coupling packaged structure (b) The image of the packaged optical microresonator

## III. PROPERTIES OF THE PACKAGED MICRORESONATORS

To measure the Q factor of the microresonator, the WGMs excitation and collection optical setup as shown in Fig. 2(a) was built. A tunable laser (NKT, Koheras BASIK Module E15) with a central wavelength of 1550 nm and a linewidth of 0.1 kHz is used as the laser source and transmitted to a tapered fiber via an optical isolator, an optical attenuator, and a polarization controller to excite the microcavities. The laser power is controlled by the attenuator to be lower than 50μw to avoid the thermal effect. The polarization controller is used to adjust the polarization state of the outgoing light. At the same time, the output beam of the fiber taper is detected by a high-speed photodetector (Newport Mode 1811). By scanning the laser frequency, the transmission spectrum of the optical microresonators can be observed in the oscilloscope (YOKOGAWA, DLM3034). The Q factors can be obtained by calculating the full width at half-maximum of the resonance dip. Comparing the Lorenz fitting spectrum before and after the packaging of the same resonant mode from one cavity as plotted in Fig. 2 (b) and (c), it can be found that our package method can maintain the mode structure well, and the Q factor of the packaged microcavity is up to $10^9$ magnitudes.

When the Q factor of the microresonator exceeds $10^8$, along with the tuning speed of the laser is increased, the ringing phenomenon of resonant mode can be observed. As shown in Fig. 4 (c), by extracting maxima and minima from the resonance, the envelope of the oscillations indicates an exponential field decay from the resonator with a time constant of 1.69 μs and 2.30 μs, corresponding to a loaded Q factor $1.03 \times 10^9$ and $1.40 \times 10^9$.

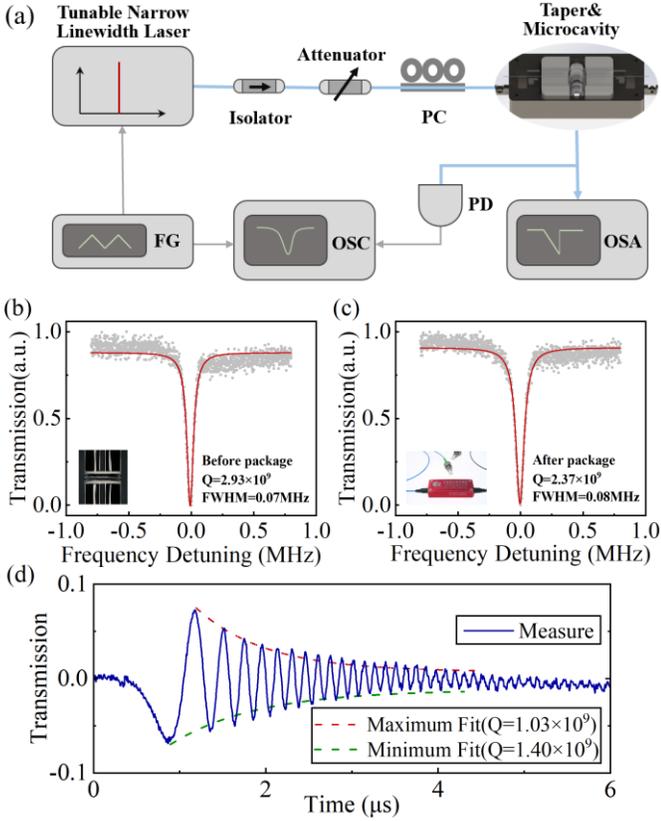

**Fig. 2.** Experimental setup and results. (a) Schematic diagram of the WGMs excitation and collection experiment setup. PC, Polarization Controller; PD, Photodetector; OSA, Optical Spectrum Analyzer; FG, Function Generator, OSC, Oscilloscope. (b) and (c) are comparison diagrams of the same typical resonant mode in the Q for unpackaged (left) and packaged (right) microresonators. The gray scatter in the diagram is the collected resonant mode waveform, and the red is the Lorentz fitting waveform. (d) Typical ringdown spectra of microcavity devices, the fitted exponential decay time ($\tau$) are 1.69 $\mu$s and 2.30 $\mu$s corresponding to a loaded Q-factor $1.03 \times 10^9$ and $1.40 \times 10^9$, respectively.

The homogeneity of the properties of packaged microcavities is depicted in Fig. 3. Five silica microresonators with diameters of 5 mm were fabricated with the same fabrication parameters on a single microrod, as shown in Fig. 3(a), and then the microresonators were packaged as shown in Fig. 3(b). The results plotted in Fig. 3 (c) and (d) show that our packaged microcavities have good consistency in performance parameters (Q factor error < 10%, FSR error < 2%).

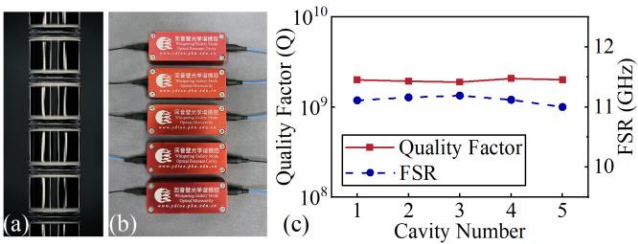

**Fig. 3.** Device homogeneity. (a) Microscope image of five silica microresonators. (b) Packaged optical microresonators image. (c) Schematic diagram of Q factors and FSR error of packaged microresonators.

## IV. ROBUST EXPERIMENTS

As the key and core components of such systems as narrow linewidth lasers, soliton light sources, and ultrasensitive optical sensing, the lifetime and stability of the performance parameters are critical for whether optical microcavities can go out of the laboratory for extensive applications. Therefore, robust experiments on packaged silica microresonators are carried out to simulate the interference of uncontrollable factors, such as hot and humid environments or mechanical vibration, which may be encountered during the routine use, transportation, and storage process, to verify the practicability of the packaged structure.

We position the packaged silica microresonators in ultraclean (class 1000) indoors with constant temperature (25°C) and humidity (35%) and outdoor blinds with uncontrollable temperature and humidity for durability tests, respectively, and the variation over time in Q factors of the microcavity were recorded simultaneously. Through two-month Q-factor-tracking experiments of the packaged structure, the characterization of durability for the optical microcavity was achieved. Figs. 4(a) and 4(b) correspond to the Q change of the same resonant mode placed indoors and outdoors, respectively. It can be seen that there is no obvious deterioration in Q factors of the packaged microcavities, even though the ambient temperature and humidity fluctuate considerably in outdoor atmosphere exposure, and the Q factors are always greater than $10^9$. This sufficiently demonstrates that such a sealed structure plays a good role in protecting the coupling system from external intrusion by virtue of outstanding air tightness, water tightness, and thermal properties.

Besides, the vibration tests of the optical microresonators were also performed. The packaged microresonators were pegged on the vibration testing apparatus, and three-axis simulation vibration with an amplitude of around 1.5 mm in the longitudinal, horizontal, and vertical directions was set up. The vibration time at each fixed vibration frequency point was set as six hours. The results plotted in Fig. 4(c) display the variation in the Q factors and normalized transmission of the optical microcavities with the vibration frequency. As shown in the diagram, the Q factors of the microresonators do not decrease significantly, and the normalized transmission can still be maintained above 90%. Additionally, a six-hour frequency-varying vibration test was performed. The results manifest that the packaged microresonators still remain a high Q factor (>$10^9$) and high transmission (>90%). Thus, the packaged structure proposed has good vibration isolation performance.

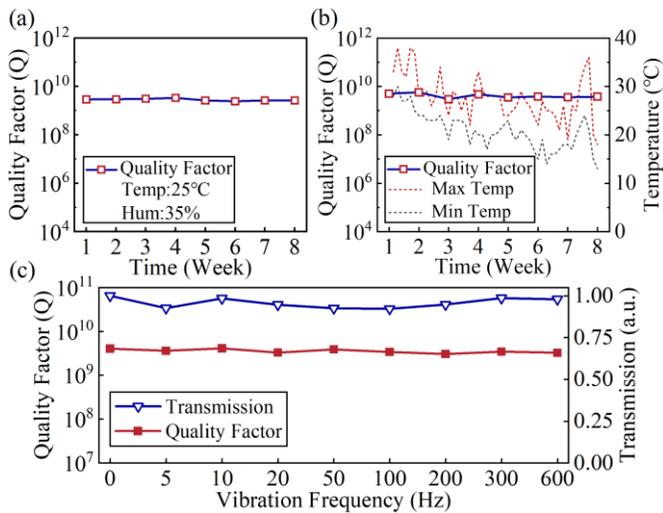

**Fig. 4.** Robust experiment. (a) and (b) are the attenuation diagrams of the Q factors of packaged microresonators placed indoors and outdoors respectively, the red and black broken lines represent the maximum and minimum temperature changes in eight weeks respectively, and the time points with high humidity have been marked in the figure. (c) Tested Q and normalized transmission of packaged microresonators after vibration experiments, the blue and red dotted lines represent the changes of Q factors and normalized transmission with the vibration frequency respectively.

## V. Conclusion

In this paper, we designed and fabricated packaged silica microresonators with compact structure, good versatility, high robustness, and other excellent features. The Q factors as high as $10^9$ were obtained with great homogeneity. In addition, the durability and vibration tests have fully confirmed that the packaging technique can maintain the device performance under normal ambient vibration of 0~600Hz, and there is no obvious degradation of high Q factors in outdoor exposure for two months. The above packaged optical microresonators are applicable to practical devices including external-cavity narrow linewidth lasers, ultra-narrow optical filters, and soliton light sources, with the applied superiority of reducing overall cost, improving system stability, and batch preparation. This study paves the way for the widespread and potential use of optical microcavities in aerospace, navigation, and vehicular systems. Moreover, it also provides a solid practice foundation and valuable reference for specialized packaging in other scenarios such as optical precision sensing and high-speed optical communication.